# LA-UR-25-21899

**Approved for public release; distribution is unlimited.**

| | |
|---|---|
| **Title:** | Determining Magnetic Field Directions and Omni-directional Fluxes of Particles from STPSat-6 ZPS Plasma Measurements |
| **Author(s):** | Chen, Yue<br>Morley, Steven Karl<br>Fernandes, Philip A.<br>Larsen, Brian Arthur |
| **Intended for:** | Report |
| **Issued:** | 2025-04-15 (rev.1) |

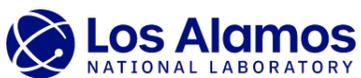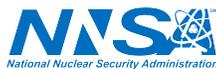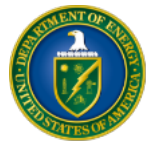

# Determining Magnetic Field Directions and Omni-directional Fluxes of Particles from STPSat-6 ZPS Plasma Measurements

**Yue Chen**[1], **Steven K., Morley**[1], **Philip Fernandes**[1], **and Brian Larsen**[1]

[1] Space Science and Applications (ISR-1) Group, Los Alamos National Laboratory

**Abstract:** This report summarizes our recent data derivation efforts on STPSat-6 ZPS plasma measurements, presenting final outcomes. We begin by outlining the methodology developed to determine local magnetic field directions from measured directional intensities of ~20 keV ions, leveraging symmetry in particles' pitch-angle distributions, and the determined field directions over a storm period are shown with errors analyzed through comparison with NOAA GOES measurements. Next, we further assess the validity of the method and quantify its overall high performance while acknowledging certain caveats. Finally, given local magnetic field directions and ZPS data, we explain how omni-directional fluxes of particles can be derived from ZPS measurements with limited directional coverage, and estimate the related errors under several theoretical scenarios. The methods developed in this study can be applied for processing and augmenting ZPS data for the next, and the insights gained here can also inform instrument design in the future.

## 1. Background

The near-Earth space radiation is often influenced by eruptive solar events, which introduce large amounts of new plasma particles, including low-energy ions and electrons (~1s-100s keV), from the magnetospheric tail region. These particles trigger space weather events inside the near-Earth environment, such as reconfigurations of the global magnetic field with enhanced current systems and sustained high levels of trapped MeV electrons [Chen et al., 2014b]. As a result, satellite systems can be adversely affected. For example, ~keV charged particles will accumulate on satellite surfaces, causing externally charging, while energetic ~MeV electrons can penetrate and internally charge dialectic components inside [Fennel et al., 2002]. Both external and internal charging often end up with sudden electrostatic discharging arcs that interfere with instrument performance or even damage entire satellite (e.g., the Galaxy 15 incident). Therefore, mitigating the effects of space weather on space assets and ensuring continuous instrument function require substantial efforts in monitoring and understanding dynamics of those space charged particles, to which Los Alamos National Laboratory (LANL) particle instruments have long contributed.

The geosynchronous (GEO) orbit is unique to space fields because satellites placed in this orbit can not only monitor the natural space environment in situ but also provide critical measurements of injected plasmas to help specify global space weather. Among other pioneers, LANL flew the first generation of charged particle instruments aboard its GEO satellites in 1976 [Higbie et al., 1979]. Since 1989, LANL has been fielding a new set of plasma and energetic particle instruments, including MPA [Bame et al., 1993], SOPA [Belian et al., 1992] and ESP [Meier et al., 1996], on a series of GEO satellites that continues to operate. Recently, LANL designed the latest generation of space particle instruments that have been deployed and tested on new GEO satellite platforms. With continuous, simultaneous multi-point measurements and decades of coverage, LANL GEO charged particle data sets have been a unique and valuable resource for scientific discoveries and operation innovations in space weather and will play a significant role in this emerging era of machine learning.

## 2. STPSat-6 Satellite and ZPS instrument

Space Test Program Satellite-6 (STPSat-6) was successfully launched on Dec. 7, 2021, into its GEO orbit. It is positioned at ~0° latitude and ~-111° longitude (Figure 1A) and carries on board the new particle instruments developed by LANL, including the ZPS plasma spectrometer.

ZPS is a new generation of plasma spectrometer designed to detect magnetospheric charged particles with energies ranging from 2 eV up to ~185 keV [Steinberg et al., 2019]. ZPS instrument contains two detector sub-components: ZPS Lo and ZPS Hi. ZPS Lo measures ions and electrons below 50 keV with 72 energy channels, while ZPS Hi measures ion and electron between ~13 keV and 185 keV with 19 channels. Both sub-components discriminate protons from heavier ions. Employing concentric hemispherical electrodes, ZPS Lo is a top-hat electrostatic analyzer whose exit aperture includes an array of five Channel Electron Multipliers (CEMs). These CEMs point to different directions, as plotted in Figure 1A, with an azimuthal field-of-view (FOV) spanning ~160°. Additionally, placed just outside the top hat entrance aperture is a pair of deflector electrodes, biased symmetrically positive and negative, to create electric fields that bend the trajectories of entering particles. In this way, the top hat FOV sweeps through 12 elevation angles within ~ ±45° for particles ≤ 20 keV, and the angle range decreases to ~ ±15° for 50 keV particles. ZPS Hi is a single pixel detector, comprised of a spherical section ESA attached with a solid-state detector that allows discrimination of particles by the energy deposited in the detector. This report focuses on the directional measurements made by ZPS Lo.

Since STPSat-6 has a roughly fixed orbital position relative to Earth and is also three-axis stabilized, we chose to translate all satellite-originated vectors into the geographic coordinate system (GCS) for narrative clarity, as shown in Figure 1. For example, the Nadir (East) vector in Figure 1A is mapped to ~68° (~ -22°) longitude in the GCS in Figure 1B, after translating with the vector's initial point from the satellite to Earth's center. Similarly, the looking directions of five CEMs are plotted in GCS in difference colors, compared to ZPS Hi direction represented by a red square symbol aligned with the CEM2 central direction. Additionally, daily magnetic field directions from the empirical quiet magnetic field model OP77 [Olson and Pfitzer, 1977] are shown as a gray closed path, with the average direction marked by a red symbol. The corresponding "magnetic equator" of this OP77 average direction is indicated by the red dashed line, from which one can estimate the observed particles' pitch angles as measured by CEMs. For instance, in Figure 1B, the 12 CEM2 directions (in blue) cover a pitch-angle range slightly smaller than $\pi/2$, with the southernmost directions sampling close to the 90° pitch angle.

In term of informativeness, directional particle fluxes full resolved across the entire 4π space provide deeper insights into the underlying physics compared to data with limited pitch angle coverage or even uni-directional particle fluxes. For example, the well-known Dessler-Parker-Sckopke formula [Dessler and Parker, 1959 and Sckopke, 1966] emphasizes the critical role of total energy content of injected particles from whole pitch-angle distributions (PADs), or at least omni-directional fluxes. Another example is the injected electrons during substorms, where their PADs at energies above 10 keV from midnight till noon in GEO are often dominated by intense pitch-angle diffusions [Åsnes et al., 2005]. Since the five CEMs of ZPS Lo measure different directions, as shown in Figure 1B, determining the local magnetic field directions could allow conversion of ZPS' directional measurements into PADs; however, the challenge here is that STPSat-6 carries no magnetometer on board and ZPS instrument only has a limited FOV. These factors motivated this study.

**3. Determining Local Magnetic Field Directions: Methodology, Results, and Error Estimates**

LANL has a long history of deriving local magnetic field directions from GEO particle measurements due to absence of direct magnetic field data. For instance, a methodology equivalent to principal axis analysis was successfully applied to MPA data [Thomsen et al., 1996], taking advantage of the nearly complete 4π coverage of MPA measurements. Another approach –identifying symmetric directions in SOPA and ESP measurements, which have 2π coverage in two different planes—was developed by exploiting the spinning satellite platforms and has been successfully applied [Chen et al., 2016].

The difficulty in our case is that ZPS covers a more limited portion of the 4π space, and STPSat-6 is three-axis stabilized. However, it is still possible to utilize the ion directional data from ZPS Lo, which often cover a pitch-angle (PA) range of ~ π/2 (e.g., as shown in Figure 1B), to test the symmetry of PADs in 3D space and determine the direction of the local magnetic field (B) with minimum fitting error. Here the PAD symmetry is expected to occur either around the 90° pitch angle or at the same pitch angle but from different azimuthal directions. In this study, we select ZPS Lo level-1 ion counts from three energy channels at ~20 keV for a preliminary test. All used counts are obtained by applying a three-point sliding window moving average to the original data, yielding a ~5 min time resolution to improve statistics. Ion species are chosen to avoid potential contamination in electron data, and the ~20 keV energies are selected because lower energies may be severely biased by spacecraft surface charging, while higher energies have much narrower directional/PA coverage. This report showcases ZPS data from early March 2022, although the results are applicable to other time periods.

Figure 2 provides one analysis example to explain the methodology, using measurements taken at the time point 1920 UT on March 4$^{th}$, 2022, when the satellite was near local noon. First, the northern hemisphere in GCS is divided into grids with an 1° longitude by 1° latitude size. Second, starting with the OP77 B direction as an initial guess, we examine the PAD symmetry for assumed B directions at the center of grids within 40° deviation angle from the OP77 direction. This 40° radius is chosen to balance computation time and search area, considering that OP77 B directions have a ~5° mean deviation and the 40° covers above 97$^{th}$ percentile from previous estimates [Chen et al., 2016]. Each given B direction—regardless of how far it deviates from the real direction— turns ZPS counts in PADs, whose symmetry around the 90° pitch angle or at same pitch angles is examined by fitting. For example, for one given B direction, the PADs for five CEMs and three energy channels are plotted in Panels A1, A2 and A3. Using the PAD of CEM5 (yellow) in Panel A1 as an example, we treat the majority data points on the right wing (i.e., with pitch angles > 90°) as the "real" PAD and interpolate them to the pitch angles of the five points on the left wing, thus get the absolute difference between the fitted counts and measured counts. By normalizing with the measured counts, we obtain an averaged relative error percentage, which is 5.26% for CEM5 in this case. This error calculation is repeated for all CEMs, yielding a weighted error percentage of 4.85% for the 18.7 keV channel in Panel A1 after examining a total of 17 data points. Here larger weights are assigned to CEMs with more checked data point and higher CEM anisotropy (defined as the maximum minus then minimum counts, normalized by the minimum). By repeating this process for all three channels, the all-combined error percentage is calculated at one grid point in the latitude/longitude space. By repeating this error calculation for all grids, we complete the **Step 1 of the method that is to calculate a global distribution of combined error percentages** shown in Panel A4**, and theoretically the true B direction corresponds to the location with the global minimum** **error in the distribution**. In this example, the global minimum error is located at 80° latitude and 44° longitude, marked by the red diamond symbol in Panel A4. This location represents the determined B direction from ZPS measurements. In addition to this global minimum at high latitude, Panel A4 also shows four blue fingers, each indicating local minimum errors. The determined B direction is also compared to OP77 and CEM directions in Panel B4.

It was discovered that using counts from one single energy channel often leads to erroneous results, as the example shown in Panel B1. For the same time point, the error distribution here is calculated from only the 21.6 keV channel for symmetry checking. While the distribution resembles the one in Panel A4, the global minimum now appears within one of the four blue fingers at mid-latitude (~54°), and the resultant PADs are shown in Panel B2. In this case, CEM5 has a small error percentage of 1.76%, but with fewer or zero data points checked for other CEMs, which explains the low total error percentage of 1.76%. However, since the peak locations of PADs from other CEMs deviate from the 90°, this determined B direction is unlikely to be the true one. Indeed, the four blue fingers—called "local minimum attractors" hereinafter—in Panel B1 can be attributed by the small numbers of data points checked in this region, as shown in Panel B3. These local minimums, caused by the low numbers of checked data points due to the limited instrument FOV, can be alleviated by using multiple channels but cannot be fully eliminated, as discussed further below.

By applying Step 1 described above, local B directions can be determined from ZPS Lo data at any given time. Figure 3 presents results for seven days between March $3^{rd}$ – $9^{th}$, 2022, which includes a moderate geomagnetic storm with the minimum Dst of ~-55 nT in the afternoon of March $5^{th}$. Due to the absence of in-situ measurements, we interpolate magnetic fields measured by GOES-16 (located at -75.2° longitude) and GOES-17 (at -137° longitude) to the STPSat-6 position for local B directions, referred to as GOES proxy directions, and treat them as the "true" directions for comparison. Panels in the left column show results after Step 1 by identifying global minimum error locations. In Panel A1, the average locations of four local minimum attractors are marked out by four black squares, which from right to left are: α located at 115° longitude and 50° latitude, β at (80°, 55°), γ at (50°, 55°) and δ at (10°, 50°). Many determined B directions cluster around these four minimum attractors. Panel B1 plots the latitudes of the determined B directions, where data points near the attractors have mid-latitudes below ~60° and are unlikely to represent the true B directions, based on above discussion.

To address this issue, we designed two more steps to reduce those mislocated data points: **In Step 2, instead of relying solely the global minimum error, we identify a list of local minimum error locations for each time point, and among them, select the one corresponding to the "true" B direction by not only comparing error values but also weighting the results based on the numbers of checked data points.** Improvements from Step 2 are clearly evident when comparing Panels in the central column of Figure 3 to those in the left: Panel A2 has much less points near the local minimum attractors, and Panel B2 shows most mid-latitude points relocated. To further reduce noises in the results, **Step 3 applies nine-point sliding window moving average to the results from Step 2,** and the final determined B directions are shown in the Panels in the right column.

These final determined B directions are replotted in Figure 4 for more detailed examination. In Panel A, the derived local B directions are plotted in GCS compared with the OP77 directions. For nightside points (with local times between $21^h$ and $03^h$), both derived and OP77 directions exhibit a similar general trend of shifting towards low latitudes (i.e., more stretched field lines) as expected. The locations of four local minimum attractors are again marked out by the four black squares, with few data points nearby.

Panel B shows the distribution of GOES proxy directions for comparison. Panel C plots the deviation angles between the derived B and GOES proxy directions, with noticeably larger deviations observed during the main and early recovery phases between March 5$^{th}$ and 6$^{th}$. Panels D and E show the latitudes and longitudes of the derived B directions. Their dynamics generally track the GOES proxy directions closely and are mostly confined within the bounds of the GOES-16 and -17 measurements.

Figure 5 further quantifies the errors statistically based on the results above. Panel A displays the occurrence probability distributions as a function of deviation angles between the determined B directions and GOES proxy directions. The red curve from Step 3 results has the highest and narrowest peak, located within [0$^o$, 10$^o$], with the smallest mean deviation value of 6.27$^o$, compared to the curves from Step 2 and Step 1 results which exhibit long tails and flattened peaks with larger mean deviation values. Panel B shows the cumulative distribution function for deviation angles. Again, the final results after Step 3 have the steepest curve, with a median value of 5.39$^o$ and 11.0$^o$ at the 90$^{th}$ percentile. These values are comparable to those from our previous study by Chen et al. [2016], where, compared to GOES-10 measurements over one year, deviation angles from the OP77 quiet model have a mean value of 5.02$^o$, a median of 2.52$^o$, and 10.1$^o$ at the 90$^{th}$ percentile, and other dynamic empirical models have even smaller values (see Figure 3 in Chen et al. [2016] for details).

**4. Validating and Assessing the Performance of the Methodology**

Fully validating and assessing the above methodology is challenging due to the absence of in-situ B data. While GOES proxy data provide an approximate, they are still not direct measurements. Our solution is to generate pseudo ZPS Lo counts by forcing PAD symmetric to address the issues circumstantially. That is, for any given time point with an arbitrary assumed "true" B direction, we first turn the original ZPS directional counts in PADs. Then, for each CEM, we treat one wing of the original PAD with majority data points as the real distribution, interpolate them to the pitch angles of the rest data points on the other wing, and replace the original counts with the interpolated values. In this way, we create a pseudo PAD which is forced to be symmetric around the assumed "true" B direction while retaining the general PAD shapes and count fluctuations in original counts. Repeat this and one can have a pseudo dataset over days, with the assumed "true" B directions. To avoid mathematically perfect solutions, we intentionally shift the "true" B directions off the latitude/longitude grid points by 0.4$^o$ in both directions. Finally, we apply the same B derivation method as described in Section 3 to this pseudo dataset to determine the local B directions, and then we compare the results to the known "true" direction to assess validation and error estimates.

Therefore, for an assumed "true" B direction, we generated pseudo ZPS counts over the same seven-day storm period and then repeat this process for a list of assumed directions. In terms of statistical significance, this approach provides a sufficient range of forced PADs with varying shapes under different geomagnetic conditions and local times for each single B direction. Figure 6 presents one such example, where the "true" B directions are assumed to be fixed at 70.4$^o$ longitude and 65.4$^o$ latitude. Panel A plots the derived local B directions in GCS, compared to OP77 directions and the cross-haired "true" B direction. Most results are clustered closely around the assumed B direction, with two branches extending toward the local minimum attractors β and γ. Latitudes (longitudes) of the derived B directions are plotted as a function of time in Panel B (C), compared to the OP77 initial guesses and the fixed longitude (latitude). Panel D plots the deviation angles between derived B directions and the assumed "true" one. The mean/median deviation angle value is 0.66$^o$/0.43$^o$, and 98.1% of data points having deviation angles <5$^o$. For the time point shown in Figure 2, the determined B direction lies on the grid point right next to the "true" B direction with a deviation angle of 0.62°. This is an excellent result, considering the 1° grid resolution and the at least 0.4° offset of the "true" B from a grid point.

Following the same procedure, we sampled the GCS longitude-latitude space with 21 assumed "true" B directions, as plotted in Figure 7. These directions span a wide region, with the central 17 located within longitudes [40°, 80°] and latitudes [60°, 80°], which cover the majority of the derived directions as shown in Figure 4A. In terms of mean deviation angles in Panel A, the central 17 directions all have small mean deviations <2.40°, while the two directions at the higher latitude (85°) have mean deviations ~5°, and the two directions at the lowest latitude (45°) have the highest mean deviations >7°. Regarding the percentages of data points with deviation angles < 5° in Panel B, the central 17 directions have high percentages, all exceeding 91%, while two high-latitude directions have ~84%, and two low-latitude directions have the lowest percentages. The non-weighted averages of the central 17 directions yield mean/median deviation angle value of 1.11° /0.45° and 96.5% of data points having deviation angles <5°. These values validate the methodology described in Section 3, demonstrating high performance within a confined latitude range, although the performance deteriorates quickly outside of the range. It is important to note that no extra noise is added when PAD symmetry being enforced to generate pseudo ZPS data, which may explain the lower deviation angles here compared to those in Section 3.

Therefore, based on the pseudo ZPS dataset with forced PAD symmetry, the above average values for the central 17 directions suggest better performance than all three empirical magnetic field models used in Chen et al. [2016]. In that work, the quiet OP77 model had a mean/median deviation value 5.02°/2.52°, with 77% of deviation angles < 5°, while the storm-time dynamic T01s model [Tsyganenko et al., 2003] had a mean/median deviation angle of 3.81°/1.53°, with 87% deviation angles < 5° when compared to GOES-10 measurements. At the same time point in Figure 2, the central 17 directions have small deviation angles within the range of [0.40°, 0.63°], while the deviations exceed 36° for the two low-altitude directions.

5. Deriving Omni-directional Fluxes: Methodology, Results, and Error Estimates

Given a magnetic field direction, one can extrapolate particle measurements over limited directions to construct the full PADs, thus maximize the usage of observations. In this preliminary analysis, we show how this can be done for ZPS Lo data, with potential errors evaluated theoretically. We adopt a theoretical normalized PAD function form of $j(PA)= \sin^n(PA)$, where PA is the pitch angle, and the exponent $n$ determines the anisotropy in the PAD. Clearly, negative exponents result in higher fluxes at ~0°/180° PA, positive exponents yield the maximum flux at 90°, and exponents close to zero give almost isotropic PADs. This simple function form[1] has been successfully used in many previous studies, with its validity demonstrated [e.g., Gannon et al., 2007]. For GEO, we also assume empty loss cones with a fixed size of 2.5°. Two major error sources are considered here: an imperfect magnetic field direction deviating from the real one and the noise levels in data due to measurement fluctuations of the instrument. Here again we seek answers by sampling through the parameter space.

Figure 8 displays three typical PAD types of ions commonly observed at GEO, and we use these examples to explain how PADs and omni-fluxes are derived, along with quantified errors. Panel A shows a cigar-shaped PAD, with maximum fluxes near the loss cones, which can be associated with Fermi acceleration of particles by depolarized magnetic field lines. In this example, the exponent $n = -2$ is used, and the "real" PAD shape is represented by the red curve with the "real" omni-directional flux having a value of 47.98. Ideally, CEM3 at 10.6 keV should measure "real" fluxes measured as the red symbols, which are mainly inside the low flux portion; however, due to a predetermined 5° offset in the given B direction and 10% noise in measurements, CEM3 has the "measured" fluxes shown as green symbols, with slightly different values at misplaced PA values. (Here the 10% noise is applied by multiplying a random factor within [0.9, 1.1] to the "real" fluxes.) By fitting the measured fluxes, a fitted PAD (the blue curve) is obtained, leading to a fitted omniflux value of 62.18. The absolute difference between the "real" and fitted omnifluxes, normalized by the real flux value, gives a 29.6% error percentage in the fitted omniflux in this example. Another method of obtaining omniflux is by averaging all measured CEM fluxes and timing the number by 4π, without assuming a PAD form at all. This gives another version of the omniflux called simple average (SA) omniflux, with a value of 25.2 and an error percentage of 47.5%. Here the SA omniflux has a higher error than the fitted omniflux because the "measured" CEM fluxes miss the high flux portion near the loss cones and ignore the trend in PAD. Panel B shows a simpler, almost isotropic PAD which is often observed in newly injected substorm electrons in the midnight sector at GEO [Åsnes et al., 2005]. In this case, the fitted omniflux has an error of 1.7%, and the SA omniflux error is 1.1%. Panel C illustrates a well-known pancake PAD, with peak flux at 90° pitch angle. This example has an exponent $n = 2$. Since CEM3 measures the high flux portion, both fitted and SA omnifluxes have small errors of 2.0% and 3.5%, respectively. These three examples demonstrate how errors in omnifluxes are determined, given an offset angle in the B direction and data noise level.

---

[1] Note this simple function form excludes some interesting PAD types, such as the "butterfly" PAD sometimes observed on the nightside of GEO. More sophisticated PAD function forms and detailed discussions can be found in references, e.g., Chen et al. [2014a].

Besides the two major error sources already mentioned, error levels in omnifluxes are also influenced by the alignment of real magnetic field direction relative to CEM directions, as well as the varying opening angles of the CEM for different energies. To address these issues, we again sample through the parameter space. Figure 9 shows how maximum errors are estimated for three energy values, based on "measured" CEM data with a 5° deviation in B direction and zero noise level. Panel A1 presents results for 1 keV ions. Each of the 21 grey curves plots the errors in fitted omnifluxes as a function of exponent $n$, using CEM3 data for each pre-determined B direction shown in Figure 7. The curve marked with black diamond symbols indicates the maximum errors among the 21 directions for CEM3, while the red curve represents the maximum errors among all five CEMs, i.e., the maximum errors in fitted omnifluxes. For comparison, the blue curve shows the maximum errors in SA omnifluxes. In Panel B1, gray symbols show the sampled CEM anisotropies at 21 different B directions as a function of exponent $n$. Red (blue) symbols mark the median (mean) anisotropy value for each exponent $n$. Here, the distributions of anisotropy in B1 are wider than observed in real ZPS data at 1 keV, and indeed CEM observed anisotropy values can be well represented by exponent values within [-2, 2]. For example, observed ZPS data for 1 keV ions have median, mean and max anisotropy values of 1.35, 1.77, and 125.9, respectively—values that are all covered by the calculated CEM anisotropies with $n$ within [-2, 2]. The same is true for other energies, as shown in Panels B2 and B3. Therefore, hereinafter we focus our discussion on errors within the same $n$ range. Panels A2 and A3 present results for 10 keV and 50 keV ions, respectively. For all three energies, it is observed that maximum errors in fitted omniflux for $n$ within [-2, 2] are below ~20-30% for positive $n$ values but much higher for negative $n$ values. In contrast, errors for SA omnifluxes are generally higher than fitted omnifluxes except for $n < -1$ at 50 keV. One trend in observed in fitting omnifluxes is that errors increase with higher energy, which can be explained the fact that CEM vertical FOV narrows with increasing energy, resulting in a smaller sampled PA range. In the extreme case of a single-looking direction, i.e., only one sample in PAD (such as the case with ZPS Hi), errors can be high with large exponent *n* values.

Figure 10 further investigates how maximum errors change with larger B deviation angles and higher noise levels in measurements. A comparison between Panels A1 and A2 reveals that errors in fitted omnifluxes increase noticeably with larger deviation angles. Moreover, significant increases in fitted omniflux errors are observed with higher noise levels, as seen in the solid vs dashed red curves in Panels C1 and C2, particularly for exponents with negative or close to zero values. This is understandable because, in the former case, CEM primarily sample the low flux portion, missing the majority of the omniflux, and in the latter case, high noise levels can severely distort an isotropic distribution in extreme instances. As shown by the solid vs dashed blue curves in Panels A2 and C2, deviated B directions and noise levels have limited effects on SA omnifluxes for obvious reasons: the former has little impact on simple average fluxes at all since PA values are not considered here, and the latter is largely averaged out due to the random nature of the noise application. From Panels C1 and C2, where fitted omniflux errors are often higher than 100%, one conclusion is that SA omniflux may provide a better and easier solution than fitted omniflux when the noise level in data is high and B direction is poorly determined. It should be noted that during the seven storm days studied above, ZPS Lo counts showed average noise levels ~20%, although at times the noise can be much higher.

## 6. Discussions

It is crucial to understand the formation of the four local minimum attractors as seen in Figures 3 and 4. In fact, the case study in Figure 2 provides a strong hint. Comparing the PADs in Panel A2 and B2, we observe that the PADs for CEM5 (in yellow) are similar in both panels, although the one in Panel B2 has a slightly narrower PA range. However, other CEMs examined in Panel A2 for symmetry are barely examined in Panel B2. This can be explained as follows: The determined B direction in Panel A4 and the one close to attractor γ identified in B1 both project CEM5's central direction to the ~90° pitch angle. In other words, both identified B directions roughly lie in the same plane perpendicular to the CEM5's central, but they project other CEM5 directions to different PAs. In addition, the B direction near attractor γ in B1 projects other CEMs almost perfectly contained within one wing of PADs, as shown in Panel B2, which results in fewer data points being checked for symmetry. These two factors together lead to a low total error value with fewer data points examined for symmetry in B2, and thus a wrong result in B1. Similarly, the β attractor has comparable effects on CEM4, and this pattern continues for the other attractors. Indeed, by increasing the deviation angle radius from 40° to larger values, we could expect to see a fifth attractor appearing for all five CEMs. In summary, these attractors primarily arise due to the limited vertical FOV, i.e. PA coverage, of the CEMs and their relative positions.

Obviously, the major issue in the B determination methodology is to differentiate the determined B directions that are located close to the local minimum attractors—not necessarily all wrong, but at least some of them. Steps 2 and 3, as described in Section 3, were designed for this purpose, and they work reasonably well. One approach might be to reduce the deviation radius of grid points for error calculation to avoid the attractor locations, but the risk is that choosing a smaller deviation radius may exclude the most interesting moments with highly stretched geomagnetic directions. Another solution is to use a more sophisticated dynamic magnetic field for initial guesses and then calculate errors with a small deviation radius. Additionally, much improved inter-CEM calibration can also help expand the PA coverage, thus alleviating the attractor effects. This can be understood by revisiting the PAD plots in Figure 2. For example, in Panel A2, if all CEM PADs overlap each other, the PA coverage would increase from <~90º individually for each CEM to >~115º altogether, providing more data points for symmetry examination. Furthermore, a wider PA coverage means the numbers of examined data points would distribute more evenly in the space and thus reduce the effects of local minimum attractors.

This current study provides a first glimpse into possible errors in extrapolating for full PADs and omnifluxes, highlighting the critical importance of low noise levels in data. While not tested here, we anticipate similar errors for electron species. Besides, the error estimates provided here are far from comprehensive. For example, errors may arise from the selected simple PAD fitting function form, and other PADs, such as the flat-hat and butterfly forms, are not accounted for in this analysis.

The ultimate challenge lies in the unknown truth—namely, the lack of B field in-situ measurements, despite the use of GOES proxy data has been used. One solution is to compare with a nearby GOES satellite carrying a magnetometer or with a LANL GEO satellite with MPA/SOPA/ESP instruments on board. It also helps if the STPSat6 satellite could be made to spin, even for a short time interval, to provide wider directional coverage in particle measurements, enabling more reliable derivation of B directions along with known omnifluxes. This would allow us to better quantify the performance of the two methods described above. Naturally, carrying a magnetometer on board would solve part of the issue, though it would still not provide the omni-directional flux measurements. Given that no magnetometer is planned in the future and no major hardware modifications are foreseen, both problems can be alleviated by increasing the FOV—e.g., increase the bending electric filed for CEMs to reach wider elevation-angles > ~+/-55º for the next ZPS instrument.

## 7. Conclusions:

This report summaries our recent data derivation efforts on STPSat-6 ZPS plasma measurements. We begin by describing the methodology used to determine local magnetic field directions by examining the symmetries in the directional distributions of ~20 keV ions. We then present the results of the determined B directions over a seven-day storm period, showing a performance comparable to the quiet OP77 empirical model, but less accurate than dynamic empirical models when using GOES proxy data as the ground truth. Additionally, we validate and assess the proposed method, demonstrating its high potential within a confined latitude range. Finally, given the local magnetic field directions and ZPS Lo fluxes, we explain how omni-directional fluxes can be derived from limited ZPS measurements and estimate potential errors across several theoretical scenarios. It is shown that minimizing data noise levels down is critical for both issues. The methods and results from this work can be applied for processing ZPS data in the next, and the insights gained will contribute to improving data derivation techniques and future instrument design.

.

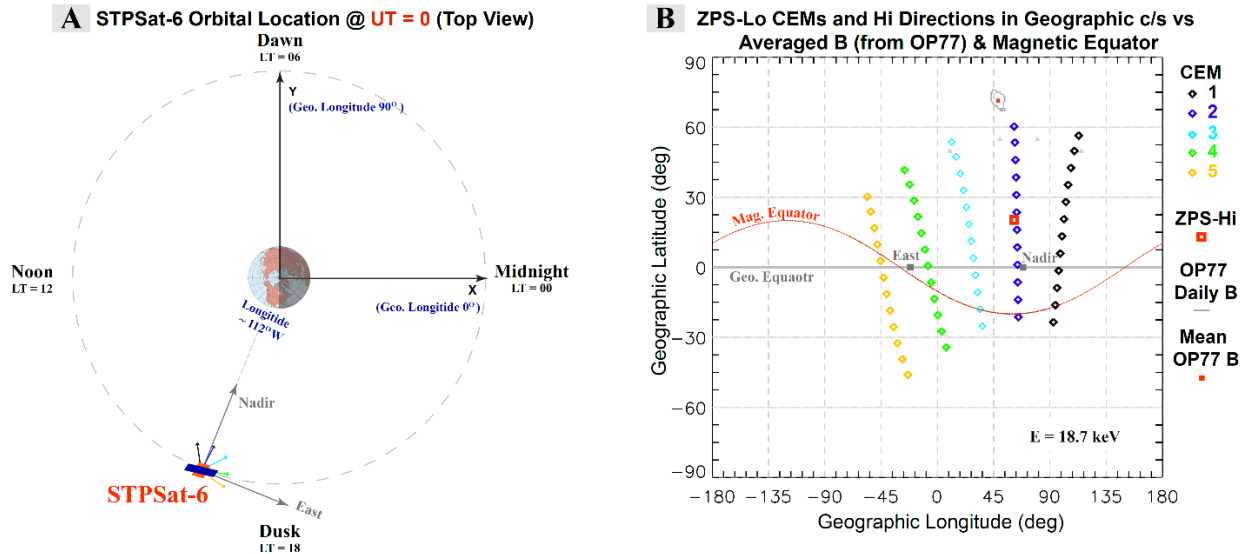

**Figure 1. The ZPS instrument on board STPSat-6 satellite measures ion and electron directional distributions along the GEO orbit. A)** Top view of STPSat-6 in GEO orbit from the North Pole at UT=0. The Sun is positioned to the left, and the X and Y axes of geographic coordinate system (GCS) are also plotted. The five short colored arrows are CEM pointing directions of ZPS Lo on board STPSat-6. **B)** Twelve looking directions of each of the five ZPS Lo CEMs (here at 18.7 keV particle energy) are mapped to GCS, along with the ZPS Hi direction (the large red square). CEM1 is in the rightmost column, and consecutively CEM5 in the leftmost. For comparison, daily magnetic field directions from the OP77 model are plotted in gray, with the averaged direction (the red symbol) at the center. The "magnetic equator" plane, derived from the OP77 mean direction, is marked by the red dotted line. The east looking and nadir directions from the satellite are also plotted in gray symbols along the geographic equator.

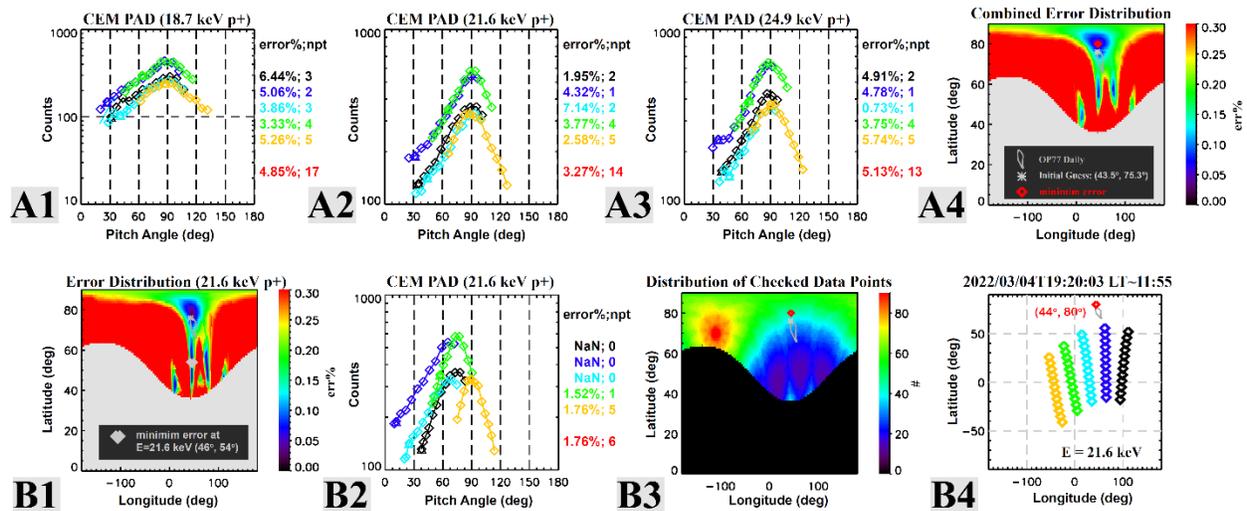

**Figure 2. An example of determining local magnetic field (B) direction from ZPS-Lo ion counts** is shown for the time 1920 UT on March 4th, 2022, when the satellite locates near local noon. **A1, A2 and A3)** Given a pre-determined B direction, PADs of five CEMs are plotted, with fitting error values presented for the three selected energy channels, respectively. **A4)** Combined fitting error distribution in GCS latitude and longitude space at 1920UT. Daily OP77 B directions are plotted in gray, with the direction at this moment marked by a gray asterisk. The location with the global minimum error is marked by a red diamond. **B1)** Similar fitting error distribution in GCS latitude and longitude space at the same time point for one single energy channel at 21.6 keV. The global minimum error location is marked by a gray diamond. **B2)** Using the determined B direction in Panel B1, PADs of five CEMs for 21.6 keV ions are plotted with fitting errors provided. **B3)** The distribution of number of fitted data points. **B4)** The looking directions of five CEMs are compared to OP77 daily directions (gray) and the determined local B direction (red diamond, as in Panel A4) in GCS.

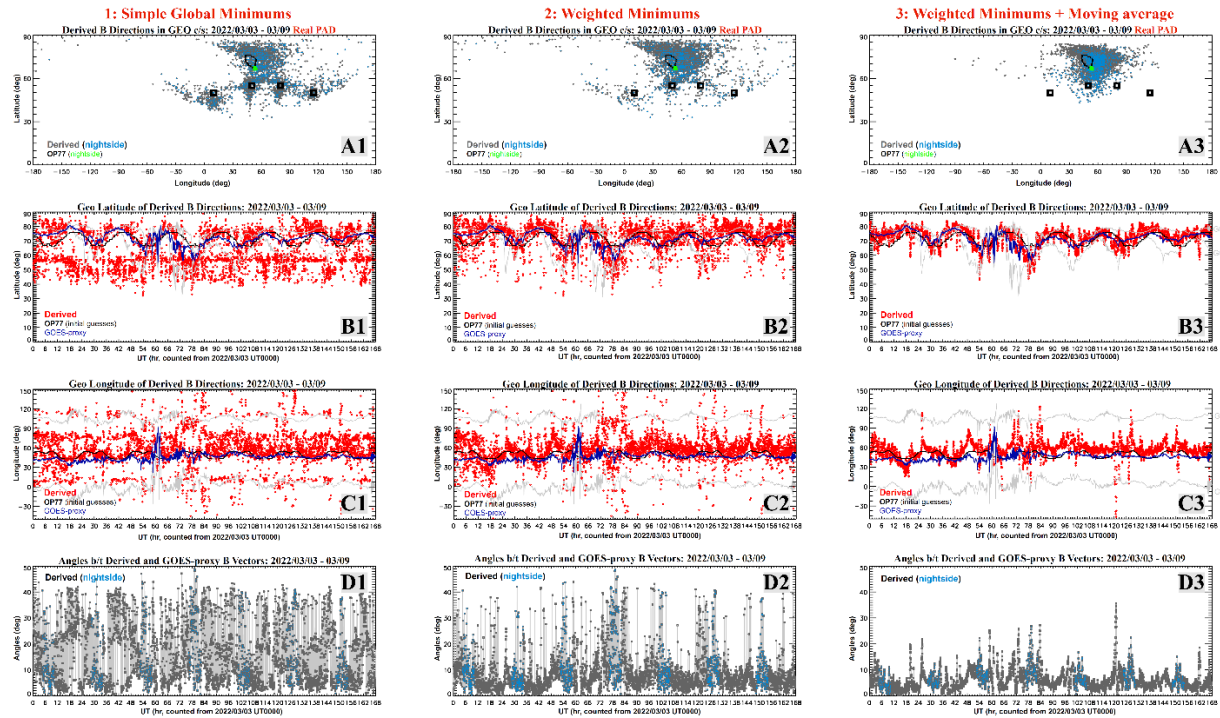

**Figure 3. Determined local B directions over seven days (March 3rd -9th, 2022) during a geomagnetic storm. The panels in the left column show results directly from the global minimum locations after Step 1. A1)** Determined B directions (gray for dayside and blue for nightside) compared to OP77 B directions (black for dayside and green for nightside). Four local minimum attractors, named α, β, γ and δ from right to left, are marked in black squares. **B1 & C1)** GCS latitudes (longitudes) of determined B directions (red) as a function of time, compared to latitudes (longitudes) from OP77 model (black) and GOES proxy directions (blue). Two gray curves represent in-situ measurements from GOES-16 and GOES-17. **D1)** Deviation angles between the final determined B directions and GOES proxy as a function of time. The panels in the central column display results from the recalculated minimum locations after Step 2, while the panels in the right column show the final results from the nine-point averages applied in Step 3.

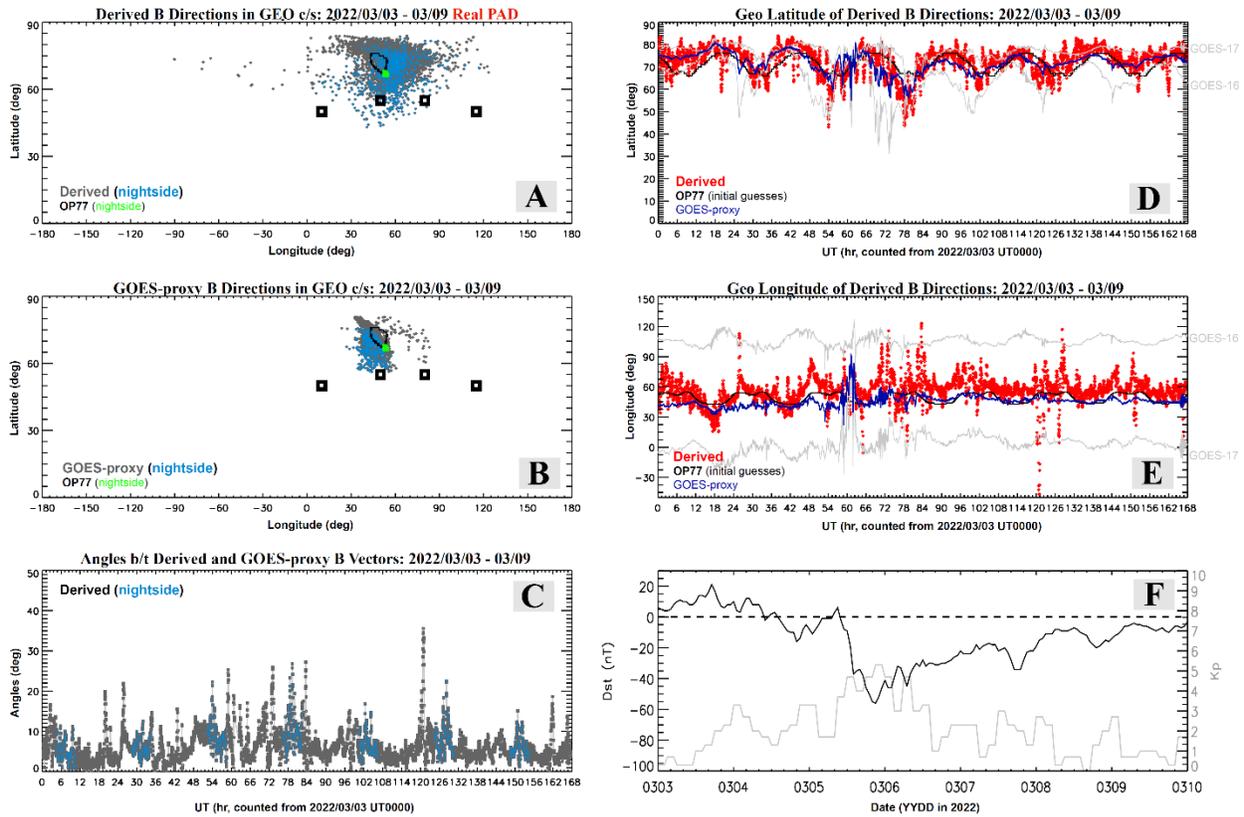

**Figure 4. Final determined local B directions over seven days (March 3rd -9th, 2022) during a geomagnetic storm, based on measured PADs. A)** Determined B directions (gray for dayside and blue for nightside) vs OP77 B directions (black for dayside and green for nightside) in GCS. The four local minimum attractors, α, β, γ and δ from right to left, are marked by black squares. **B)** GOES proxy directions vs OP77 directions in GCS. **C)** Deviation angles between the final determined B directions and GOSE proxy directions as a function of time. **D and E)** GCS latitudes (longitudes) of the determined B directions (red) as a function of time, compared to latitudes (longitudes) from the OP77 model (black) and GOES proxy data (blue). Gray curves represent in-situ measurements from GOES-16 and GOES-17. **F)** Dst (black) and Kp (gray) indices.

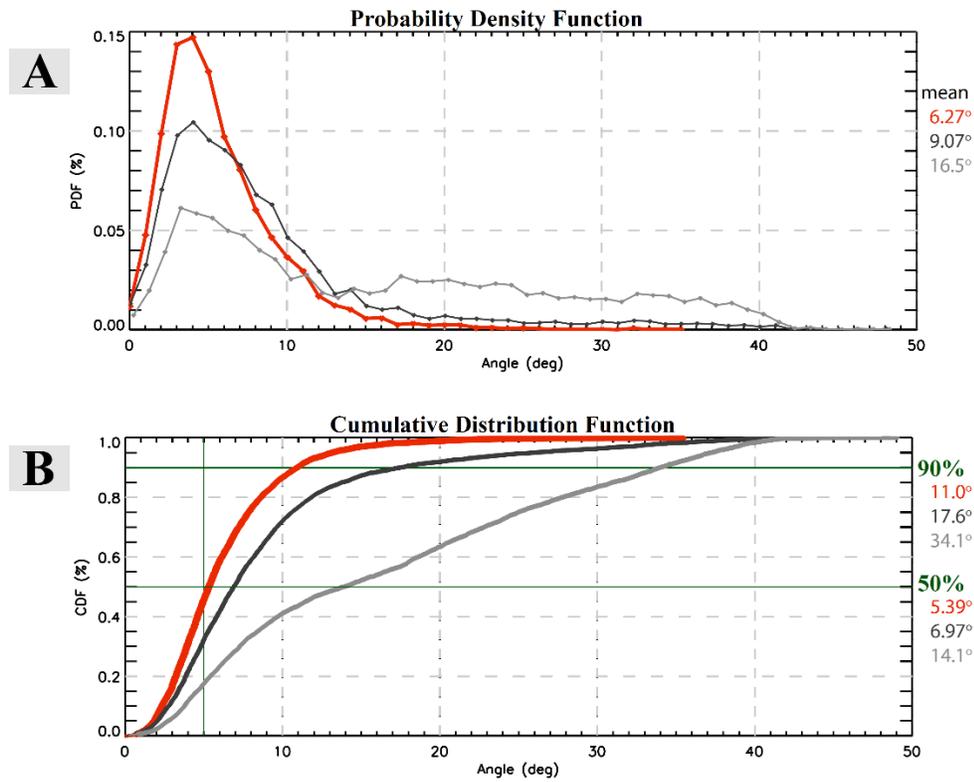

**Figure 5. Statistical distributions of deviation angles between the determined magnetic field directions and GOEX proxy directions. A)** Normalized occurrence percentages as a function of deviation angles. The red curve represents the final determination results after all three steps, the black curve represents results after Step 2, and the gray curve shows results after Step 1. Mean deviation values are provided for each case in different colors. **B)** Cumulative percentages as a function of deviation angles for the same three cases. The deviation angle values for the 90th and 50th percentiles are also provided.

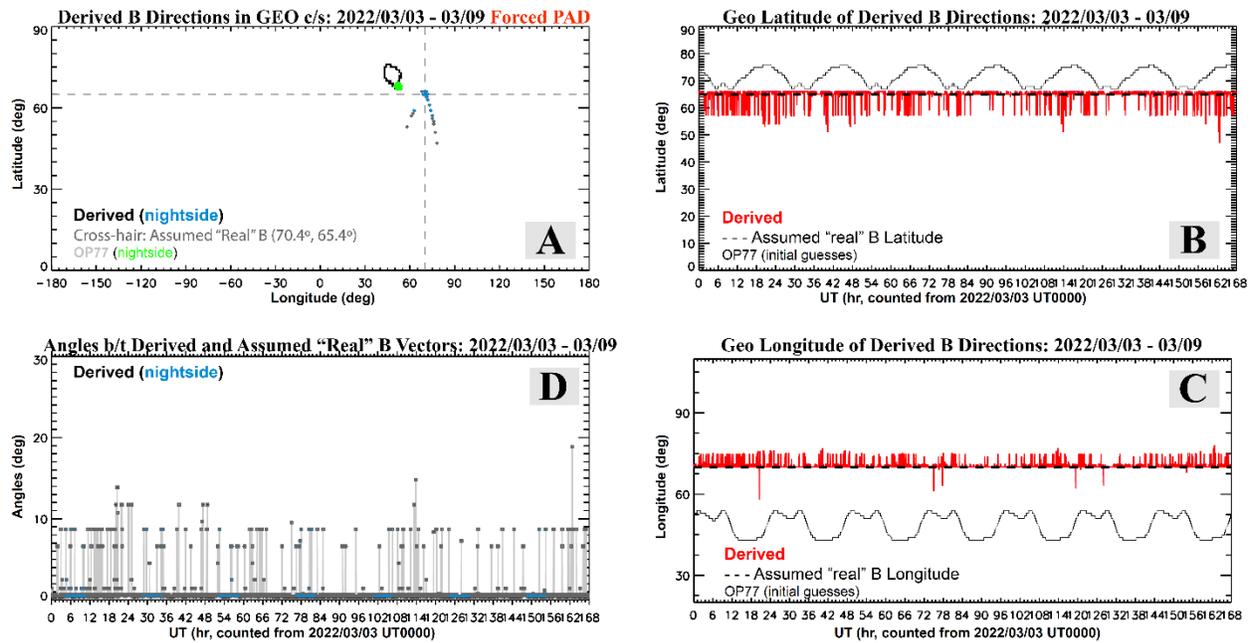

**Figure 6.** Determined local B directions over the same seven storm days (March 3$^{rd}$ -9$^{th}$, 2022). "Forced" PADs (symmetrized pseudo PADs) are used, with the assumed "true" magnetic direction fixed at (longitude :70.4$^o$, latitude: 65.4$^o$) throughout the time period. **A)** Determined B directions (gray for the dayside and blue for the nightside) vs OP77 B directions. **B)** GCS latitudes of determined B directions (red) as a function of time, compared to the latitudes from OP77 model (black) and the fixed "true" latitude (65.4$^o$) in the dashed line. **C)** GCS longitudes of determined B directions (red) as a function of time, compared to the longitudes from OP77 model (black) and the fixed "true" longitude (70.4$^o$). **D)** Deviation angles between the determined B directions and the "true" B direction as a function of time.

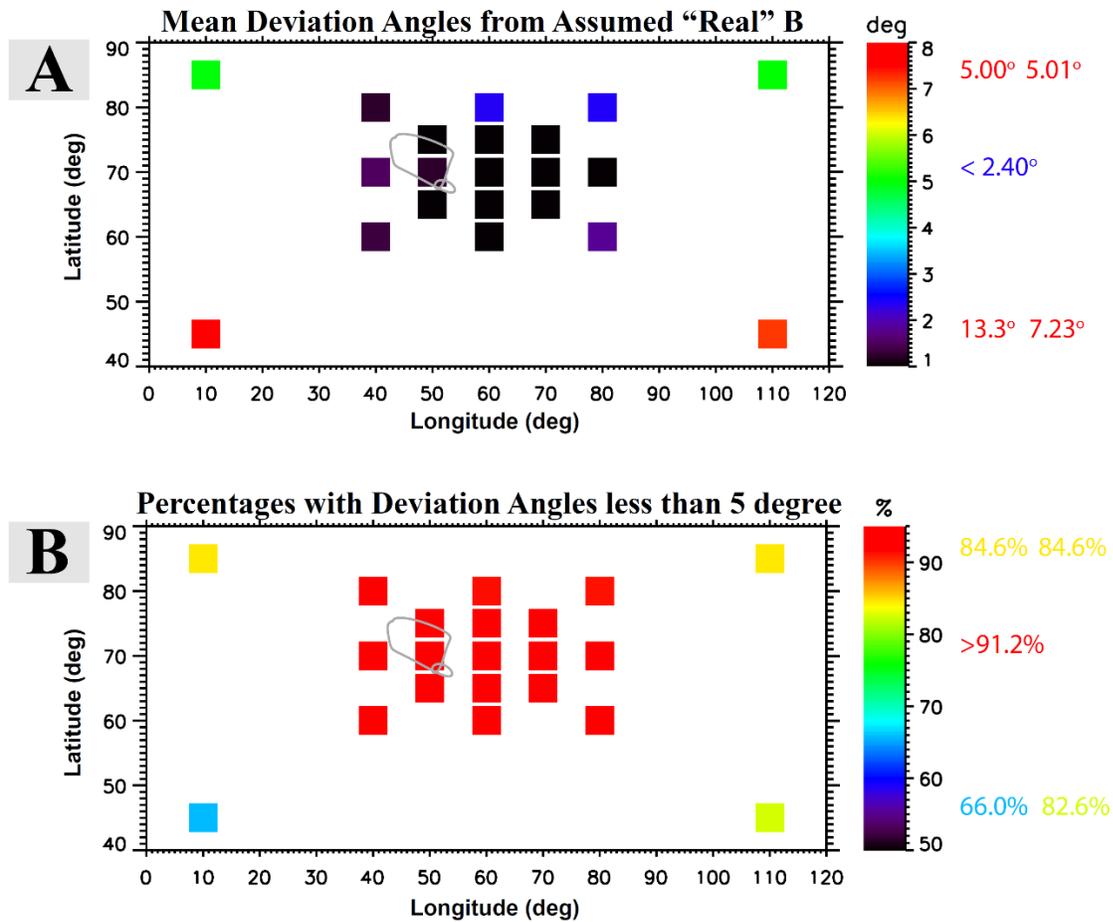

**Figure 7. Calculated errors for 21 assumed "true" magnetic field directions.** Longitudes and latitudes of the 21 directions can be read from both panels. **A)** Average deviation angles between the determined B direction and assumed "true" B directions. OP77 directions are plotted in solid gray for comparison. Center of each color block location represents the assumed B direction, with the color indicating the deviation value. Mean deviation values (or range) are provided on the right-side. **B)** Percentages of data points with deviation angles < 5º for the 21 assumed B directions. Percentage values (or range) are provided on the right-side.

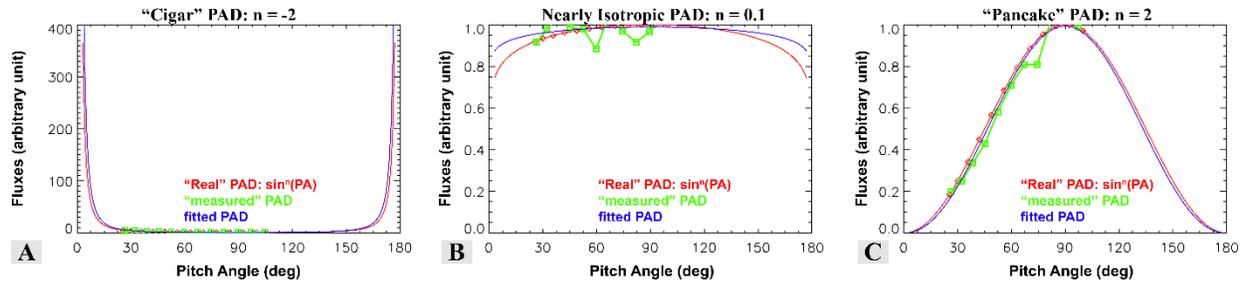

**Figure 8. Example pitch-angle distributions (PADs) commonly observed for ions at GEO, with errors introduced by imperfect B directions and measurement noises. A)** A cigar-shaped PAD has the maximum flux at B field-aligned directions. Here the sine function has a negative exponent value of -2. **B**) Nearly isotropic PAD with an exponent value of 0.1. **C**) A flattop distribution with the maximum at 90° pitch angle. Here the positive exponent has a value of 2. In all examples, the assumed true B direction points at 60° longitude and 70° latitude, and CEM3 directions at 10.6 keV are used. In each panel, the red curve plots the "true" PAD j(PA)= $\sin^n$(PA), where *n* is the predetermined exponent. The red data symbols are the "real" flux values expected for CEM3 at twelve directions, the green symbols are the "measured" fluxes by CEM3 with a 5° offset in the magnetic field direction and 10% random noise, and the blue curve represents the "fitted" PAD based upon the "measured" values.

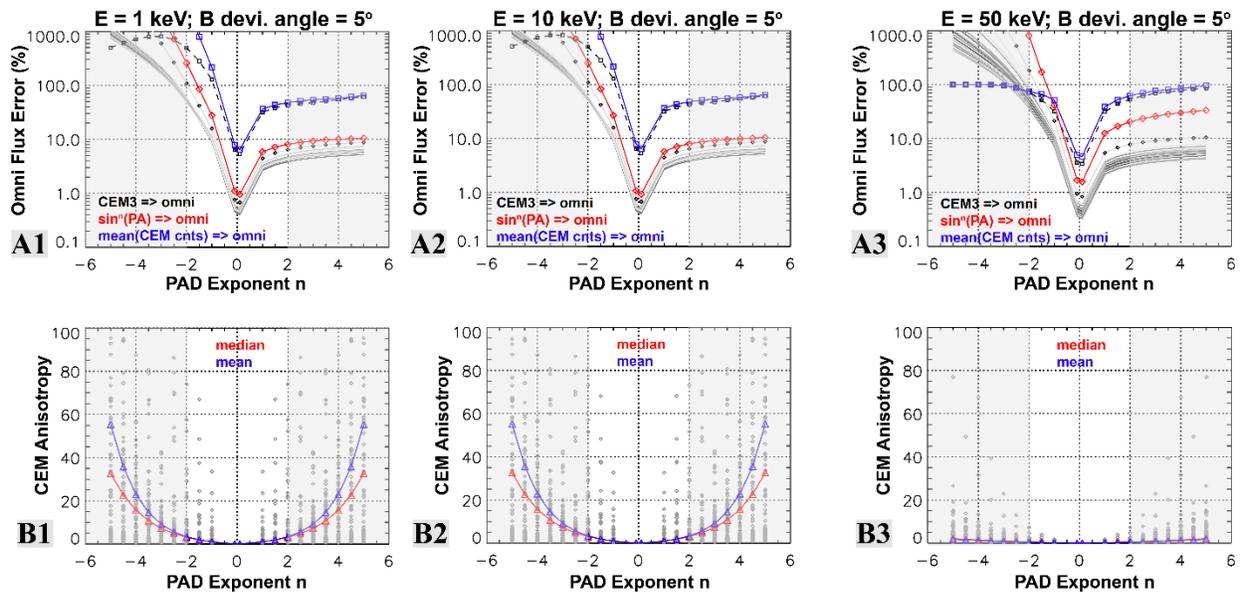

**Figure 9. Maximum errors in fitted omni-directional fluxes as a function of exponent *n*, sampled around the 21 "true" magnetic field directions, with a 5° deviation angle in each B field direction. A1**) Errors for 1 keV ions. Each of the 21 grey curve plots fitting errors in CEM3 as a function of exponent *n* for one pre-determined B direction. The curve with black diamond symbols marks the maximum errors among the 21 directions for CEM3, while the red curve indicates the maximum errors for all five CEMs. The blue curve plots the maximum errors in omni-directional fluxes by simply averaging the 12 measured counts from each CEM without assuming a B direction. **B1)** Gray symbols represent the anisotropies in measured CEM counts at all sampled cases as a function of exponent *n*. Red (blue) symbols indicate the median (mean) anisotropy value for each n. **A2 and B2)** are for 10 keV ions in the same format, and **A3 and B3)** are for 50 keV ions.

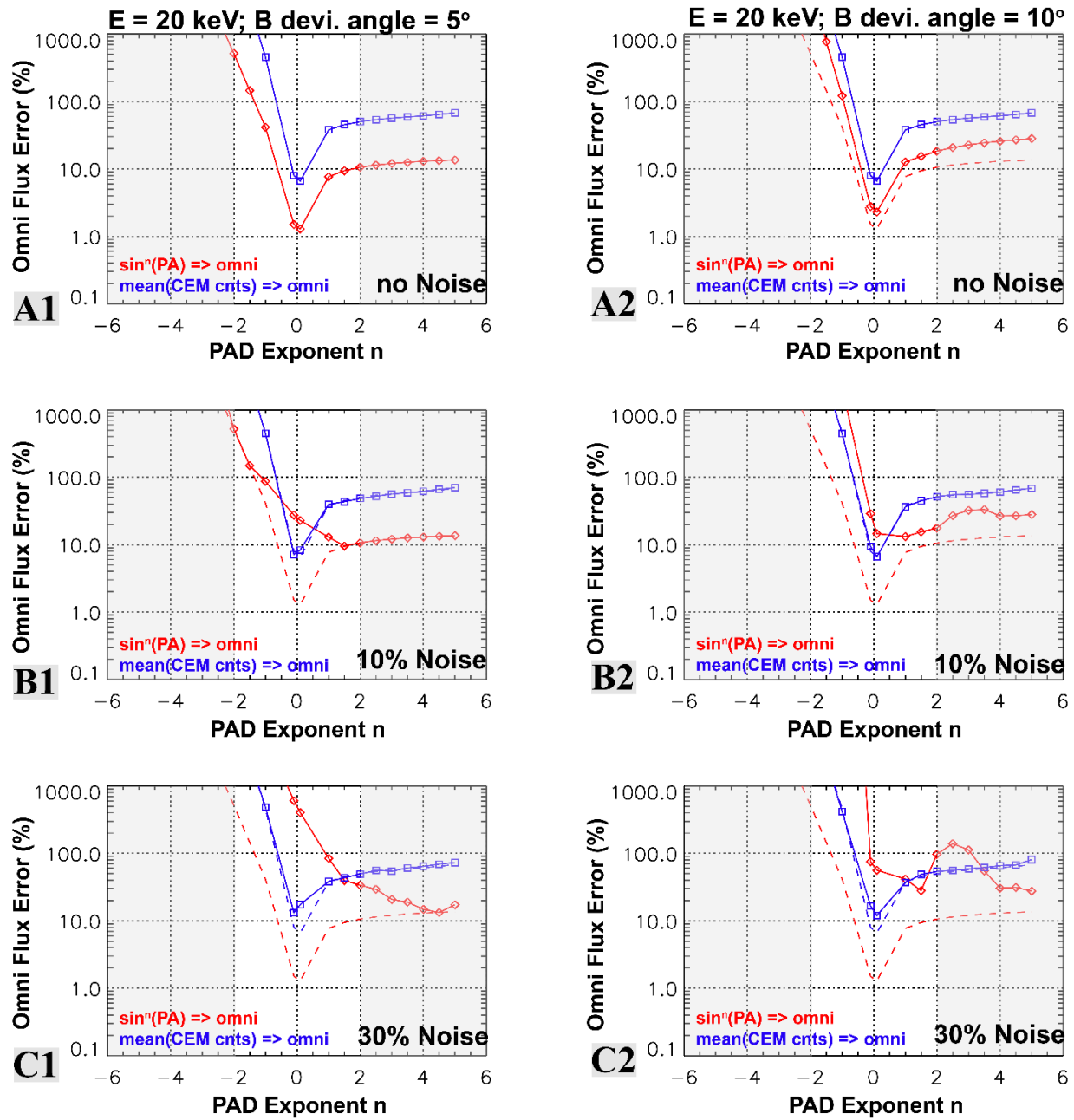

**Figure 10. Maximum errors in omni-directional fluxes for 20 keV particles at different deviation angles in B direction and noise levels.** The panels in the left column represent cases with a 5° deviation angle, and the right column for a 10° angle. From top row to bottom, the noise levels in CEM measurements increase from zero, to 10%, and finally 30%. The error curves in Panel A1 are copied to other panels in dashed lines for comparison. Red curves represent error percentages in fitted omnifluxes, and blue curves for SA omnifluxes.